\documentclass[10pt,letterpaper]{article}

\usepackage{ccn}
\usepackage{pslatex}
\usepackage{apacite}
\usepackage{natbib}

\usepackage{tablefootnote}
\usepackage{tabularx}
\usepackage{latexsym}
\usepackage{booktabs}
\usepackage{graphicx}
\usepackage{color}
\usepackage{wasysym}
\usepackage{lineno}
\usepackage{tikz}
\usetikzlibrary{shapes,arrows,positioning,calc}
\usepackage[utf8]{inputenc}
\usepackage[english]{babel}
\usepackage{caption, subcaption}
\usepackage{tcolorbox}
\usepackage{hyperref}   

\definecolor{barMainCol}{RGB}{0, 48, 82}
\definecolor{barCSLfood}{RGB}{112, 184, 93}
\definecolor{barCSLdishes}{RGB}{40, 125, 120}
\definecolor{barIAIfood}{RGB}{28, 155, 216}
\definecolor{barIAIdishes}{RGB}{0, 100, 163}
\definecolor{situationBack}{RGB}{84, 208, 255}
\definecolor{actionBack}{RGB}{180, 230, 250}
\definecolor{motionBack}{RGB}{220, 245, 255}
\definecolor{accentred}{RGB}{231, 76, 60}
\definecolor{accentyellow}{RGB}{241, 230, 100}
\definecolor{accentorange}{RGB}{253, 143, 0}
\definecolor{accentgreen}{RGB}{112, 184, 93}
\definecolor{lightblue}{RGB}{28, 155, 216}
\definecolor{darkblue}{RGB}{0, 48, 82}
\definecolor{easeblue}{HTML}{144F78}
\definecolor{outer}{HTML}{0d2f54}
\definecolor{inner}{HTML}{bee4fb}
\definecolor{light}{HTML}{e2f3fe}

\pgfdeclarelayer{background}
\pgfdeclarelayer{foreground}
\pgfsetlayers{background,main,foreground}

\newsavebox\MyPicture
\NewDocumentCommand{\roundedpicture}%
      {O{width=0.6\linewidth}
       O{draw=white!80!easeblue,line width=0.5pt,rounded corners=8pt}
       m}{%
   \savebox\MyPicture{\includegraphics[#1]{#3}}%
   \begin{tikzpicture}%
    \draw [path picture={%
                   \node[anchor=center] at (path picture bounding box.center) {%
                       \usebox\MyPicture};},#2]
          (0,0)  rectangle (\wd\MyPicture,\ht\MyPicture);
   \end{tikzpicture}%
}

\title{Neuronal Correlates of Semantic Event Classes during Presentation of Complex Naturalistic Stimuli: Anatomical Patterns, Context-Sensitivity,
and Potential Impact on shared Human-Robot Ontologies}

\author{
Florian Ahrens,$^{1*}$ Mihai Pomarlan,$^{2}$ Daniel Be{\ss}ler,$^{2}$ Michael Beetz,$^{2}$ Manfred Herrmann$^{1}$\\
\normalsize{$^{1}$University of Bremen, Department of Neuropsychology and Behavioral Neurobiology}\\
\normalsize{$^{2}$University of Bremen, Institute for Artificial Intelligence}\\
\normalsize{$^\ast$Corresponding author email: fahrens@uni-bremen.de}
}

\begin{document}

\maketitle


\section{Abstract}
{
\bf
The present study forms part of a research project that aims to develop cognition-enabled robotic agents with environmental interaction capabilities close to human proficiency. This approach is based on human-derived neuronal data in combination with a shared ontology to enable robots to learn from human experiences. 
To gain further insight into the relation between human neuronal activity patterns and ontological classes, we introduced General Linear Model (GLM) analyses on fMRI data of participants who were presented with complex naturalistic video stimuli comparable to the robot tasks. We modeled four event classes (pick, place, fetch and deliver) attached to different environmental and object-related context and employed a Representational Similarity Analysis (RSA) on associated brain activity patterns as a starting point for an automatic hierarchical clustering. Based on the default values for the Hemodynamic Response Function (HRF), the activity patterns were reliably grouped according to their parent classes of object interaction and navigation. Although fetch and deliver events were also distinguished by neuronal patterns, pick and place events demonstrated higher ambiguity with respect to neuronal activation patterns. Introducing a shorter HRF time-to-peak leads to a more reliable grouping of all four semantic classes, despite contextual factors. These data might give novel insights into the neuronal representation of complex stimuli and may enable further research in ontology validation in cognition-enabled robotics.
}


\section{Introduction}

The Collaborative Research Center for Everyday Science and Engineering (EASE CRC) aims at the development of autonomous robotic agents based on the principles of cognition-enabled robotic control through interconnected systems for self-reflected reasoning and planning that rely on a central knowledge base \citep*{beetz_cognition-enabled_2012,beetz_know_2018}. 
This knowledge base is populated by the robots’ prior experiences as well as interactions of robot and human actors in real- and virtual environments. 
The present research focuses on the intersection of human action science, artificial intelligence, and cognitive robotics, and on the conception and validation of computational models of human activity that enable a replication in robotic systems. This interdisciplinary approach fosters the development of robotic control systems rooted in cognitive principles, as well as feeding back results from robotics into human action science. 
As a technique to formalize knowledge of activities, we employ the SOMA \citep*{bessler21soma} ontology that represents tasks, situations/environments, and actions relevant to both human and robot everyday activity. This ontology thereby acts as a shared conceptualization of activities and characteristics from both the human and robot perspective.

\begin{figure}[h]
    \centering
    \input{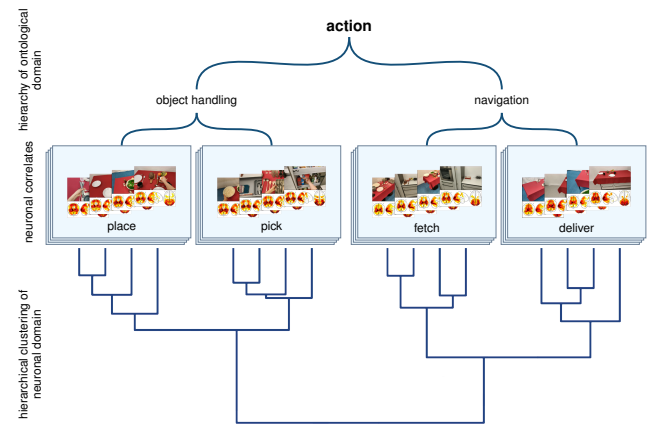}
    \caption{Neuronal correlates of semantic event classes are automatically clustered to assess neuronal validity of ontological definitions.}
    \label{fig:abstract}
\end{figure}

In the present paper, we focus on neuronal correlates of the
semantic event classes described in the SOMA ontology, as presented in \autoref{fig:abstract}. 

Specifically, by analyzing fMRI data of participants who observed first-person-perspective videos of everyday tasks, we analyzed how human neuronal activity correlates with ontological classifications.
We note that the ontology we use was created by roboticists, initially with the goal of formally defining entities that are relevant for robotic agents, and with the secondary goal of also providing a vocabulary with which to analyze human activity. Here we investigated the question as to whether the concepts and distinctions made by the ontology developers reflect distinct activity patterns, mirrored in the human brain. 

As part of the human-derived data sets, neuroimaging data of the human brain experiencing common human- and robot environmental interaction aids in shaping SOMA to correctly represent activities from the human perspective by incorporating principles of human neuronal information processing. Therefore, it seems crucial that human neuroimaging represents the activity of the human brain during environmental interaction in the context of everyday activities.
These interactions involve a high degree of active real-world engagement. In contrast, participants in neuroimaging studies lie stationary while their brain activity is measured, and only a limited degree of environmental interaction can be realized. To compensate for this drawback, the ability of the human brain to simulate environmental interaction is utilized during EASE 
neuroimaging sessions, which is hypothesized to lead to comparable brain activity patterns as found in real-world interactions. 
Through motor imagery \citep{jeannerod_representing_1994,kosslyn_seeing_1987}, purely imagined actions follow similar psychophysical rules as directly executed actions \citep{bakker_motor_2007, frak_orientation_2001} and lead to similar brain activation patterns in perceptual \citep{bird_establishing_2010,dentico_reversal_2014,dijkstra_distinct_2017,pearson_human_2019} and motor-related brain networks \citep{gerardin_partially_2000,roth_possible_1996,stippich_somatotopic_2002}. This process can be strengthened by presenting observable actions, which is hypothesized to map the observed action into the observer’s own motor system through recruitment of the human equivalent of the mirror neuron system \citep{hyeonjin_neurons_2018,molenberghs_brain_2012,rizzolatti_neurophysiological_2001}.
The activity of the neuronal network that is thought to be the basis of both motor imagery and action observation for learning, refinement, and maintenance of motor actions \citep{gallese_mirror_1998,hesslow_conscious_2002,jeannerod_neural_2001} allows a close approximation of neuronal correlates of real-world interaction when measured through fMRI, especially if stimuli are presented with a high degree of immersion \citep{caggiano_view-based_2011,jeannerod_representing_1994} and biomechanical realism through correspondence to the observing participant's motor capabilities and self-image \citep{buccino_neural_2004,stevens_new_2000,tai_human_2004}.
Through reliable activation of this network, a close approximation of neuronal correlates of real-world environmental interaction is expected possible when measuring participants' brain activity despite lack of physical engagement in the presented actions during fMRI sessions.

Actions of humans and robots are carried out in a variety of environmental contexts but also concerning task-specific requirements. Actions are performed in widely different ways and embedded in a stream of competing stimuli based on the environment and the properties of the objects on which they are performed. 

To analyze neuronal correlates of these stimuli, traditional paradigms of fMRI study-design, consisting of simple, static stimuli, often presented in a block-wise fashion, are not appropriate due to lacking of immersion and biological validity. Furthermore, signal changes in the brain measured through fMRI are primarily mediated by changes in blood oxygenation level, with a delayed increase in oxygenation after the onset of increased neuronal activity \citep{afonso_neural_2007,malonek_interactions_1996,ogawa_oxygenation_1990}. The corresponding increase in signal strength is modeled via a hemodynamic response function (HRF) 
for fMRI analysis such as the General Linear Model (GLM). This model employs linear regression to explain the signal strength of a brain voxel by variables such as stimulus classes \citep{beckmann_general_2003,poline_general_2012} and needs a prior convolution of stimulus timings with this HRF to correctly predict the underlying brain activity.
These signals thus carry a delay which can be detrimental to analysis of complex, naturalistic stimuli that exhibit various temporal characteristics, depending on the semantic entity. To assess the underlying neuronal correlates of signal changes, the HRF must correctly model the coupling between neuronal activity and oxygenation level. It was shown that HRF attributes such as time-to-peak (HRF-TTP) exhibit a considerable level of inter-subject and inter-area variability \citep{monti_statistical_2011}.

\subsection{Prior work}
Despite the limitations discussed above, researchers have begun to incorporate more complex stimuli, mainly in the form of video streams \citep{eickhoff_towards_2020,leopold_studying_2020,simony_analysis_2020},
to allow for the measurement of brain activity in a more natural and thus ecologically valid state. 
Analysis of brain activity related to complex video stimuli has revealed insights into perception of events and event boundaries \citep{zacks_human_2001} with a high degree of in inter-participant spatio-temporal correlation \citep{byrge_video-evoked_2021,hasson_intersubject_2004}. fMRI-derived brain activity data was further used to accurately predict semantic classes within the presented videos \citep{huth_decoding_2016}. 
However, these studies were often restricted to temporally larger grained episodes. 
In contrast, the ability to accurately decipher smaller grained semantic event classes within similar streams is crucial, considering the limited duration of each instance of an event class.

To test the validity of fMRI for complex and naturalistic stimuli with variable temporal characteristics, a recording and analysis of neuronal dynamics during watching videos of everyday manipulation tasks was performed where event-dependent allocation of brain activity were detected via General Linear Model (GLM) and Independent Component Analysis (ICA) \citep{ahrens_neuronal_2021}.
This procedure involved an exploratory K-means clustering of activity patterns, although restricted to a smaller subset of contexts and a low statistical threshold for analysis of spatial activation patterns. We later introduced the same dataset to analyze correlations between ontological entities and brain signals derived from an ICA and reported a group of components that exhibited stable correlations to subsets of ontological entities with apparent contextual dominance \citep{ahrens_towards_2023}. We expected this methodological procedure to allow insight into neuronal validity of a shared human-robot ontology to confirm the suitability of approaching complex naturalistic stimuli via fMRI. 
As a limitation, stable correlations were present only in a subset of participants and, while involvement of motor areas could be demonstrated, major components were primarily associated with perceptual processing.

While context- and subject-dependent differences in brain patterns can provide insight into human information processing, a stable, generalizable neuronal signature of semantic event classes despite changing contextual factors would further validate the ontological definitions and enable the larger goal of extracting and classifying instances of narrative-enabled episodic memories based on their semantic identity as defined by SOMA ontology. This procedure would be substantiated if corresponding generalized neuronal patterns were found to encompass networks comprising brain areas hypothesized to be involved in active real-world engagement with their respective event class. 

We introduced a GLM study on the same dataset as \cite{ahrens_neuronal_2021} and \cite{ahrens_towards_2023} to analyze and cluster group-level brain activity related to selected semantic event classes with the inclusion of additional contextual factors.
We hypothesized that distinct patterns of human brain activity could be organized into predefined semantic classes, which are shaped by contextual factors. 
Due to expected changes in neuronal patterns from modeling of the HRF, we further included analysis of HRF-related values for optimal differentiation between these classes. By fine-tuning the HRF settings, we aimed to enhance the temporal alignment of these stimuli, allowing insights into complex neuronal representations.

\section{Methods}
\subsection{Stimuli}

\begin{figure}
\centering
\begin{subfigure}{0.5\columnwidth}
\centering
\fbox{\includegraphics[width=.9\columnwidth]{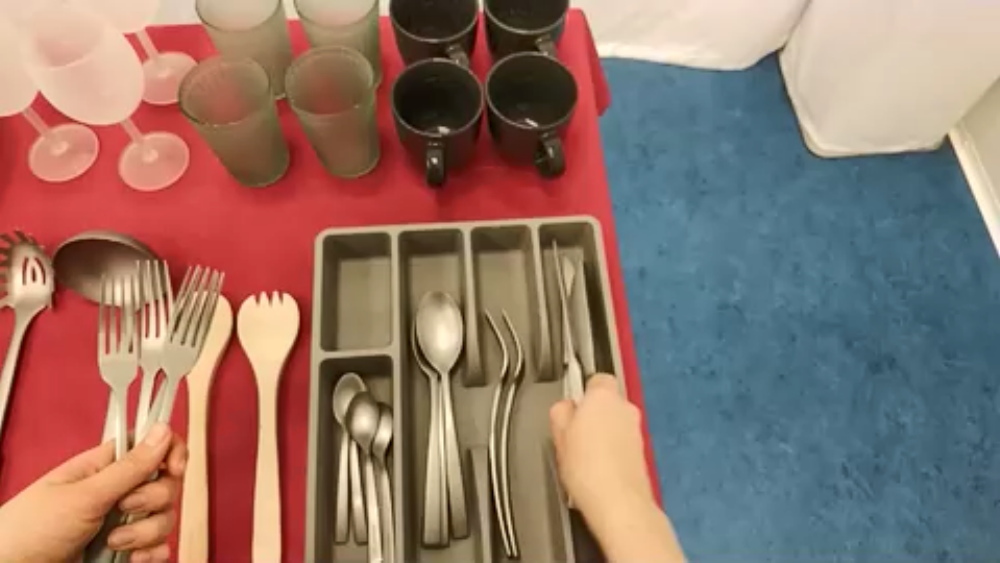}}
\caption{lab 1}
\label{fig:scr_lab1}
\end{subfigure}%
\hfil
\begin{subfigure}{0.5\columnwidth}
\centering
\fbox{\includegraphics[width=.9\columnwidth]{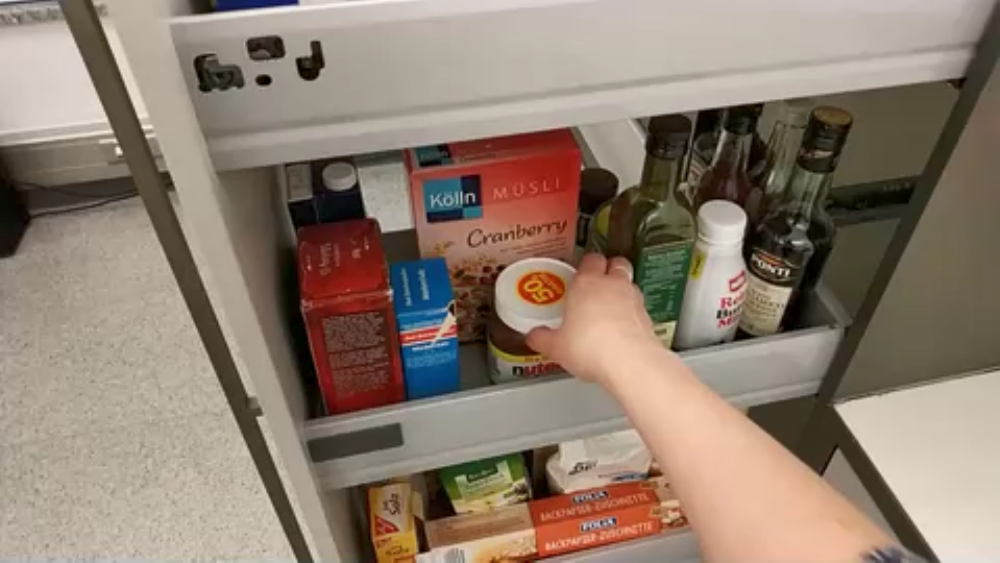}}
\caption{lab 2}
\label{fig:scr_lab2}
\end{subfigure}%
\caption{Screenshots from prerecorded 1st-person videos depicting table-setting activities in labs 1 \& 2}
\label{fig:methods_labs}
\end{figure}

Stimulus material consisted of 1st-person videos of table setting activities, recorded via head-mounted camera at University of Bremen laboratories in two different environments (\autoref{fig:methods_labs}): the Cognitive Systems Lab (lab 1) and the Institute for Artificial Intelligence (lab 2). 
The videos showed picking, transporting, and placing objects in a table setting scenario, featuring both single- and two-handed movements. Lab 1 had a simple setup with source and target tables, while lab 2 featured a modern kitchen environment. 
Videos were divided into pairs defined by item context (dishes \& cutlery versus food \& drink items). This procedure resulted in four video contexts, depending on environment and item class.
Ten videos (29-105s duration) were recorded (six in lab 1) and annotated using EASELAN~\citep{meier_comparative_2019}, a modification of ELAN annotation software~\citep{ELAN}. 

\begin{figure}
    \centering
    \resizebox{8cm}{!}{
        \begin{tikzpicture}
        [
        level 1/.style={draw=easeblue, sibling distance=6cm, edge from parent path =
        {(\tikzparentnode.south) .. controls +(0,-1) and +(0,1)
                                   .. (\tikzchildnode.north)}},
		level 2/.style={sibling distance=3cm, edge from parent path =
        {(\tikzparentnode.south) .. controls +(0,-1) and +(0,1)
                                   .. (\tikzchildnode.north)}},
        level 3/.style={sibling distance=.5cm, edge from parent path =
        {(\tikzparentnode.south) .. controls +(0,-1) and +(0,1)
                                   .. (\tikzchildnode.east)}}
        ]
          \node {\small action}
            child{node {\small object handling}
                child {node {\small pick}
                    child {node[rotate=90, anchor= east] {\small pick-dishes-lab1}}
                    child {node[rotate=90, anchor= east] {\small pick-dishes-lab2}}
                    child {node[rotate=90, anchor= east] {\small pick-food-lab1}}
                    child {node[rotate=90, anchor= east] {\small pick-food-lab2}}
                    }
                child {node {\small place}
                    child {node[rotate=90, anchor= east] {\small place-dishes-lab1}}
                    child {node[rotate=90, anchor= east] {\small place-dishes-lab2}}
                    child {node[rotate=90, anchor= east] {\small place-food-lab1}}
                    child {node[rotate=90, anchor= east] {\small place-food-lab2}}
                }
            }
            child {node {\small navigation}
                child {node {\small fetch}
                    child {node[rotate=90, anchor= east] {\small fetch-dishes-lab1}}
                    child {node[rotate=90, anchor= east] {\small fetch-dishes-lab2}}
                    child {node[rotate=90, anchor= east] {\small fetch-food-lab1}}
                    child {node[rotate=90, anchor= east] {\small fetch-food-lab2}}
                }
                child {node {\small deliver}
                    child {node[rotate=90, anchor= east] {\small deliver-dishes-lab1}}
                    child {node[rotate=90, anchor= east] {\small deliver-dishes-lab2}}
                    child {node[rotate=90, anchor= east] {\small deliver-food-lab1}}
                    child {node[rotate=90, anchor= east] {\small deliver-food-lab2}}  
                }
            };
        \end{tikzpicture}
        }
    \caption{Hierarchical event classification based on ontology}
    	  \label{fig:event_classes}
\end{figure}
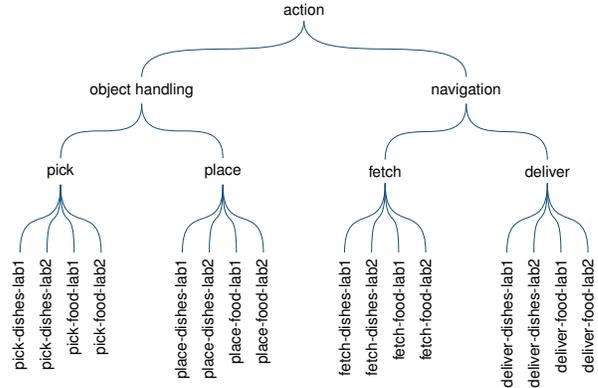

Four event classes (\autoref{fig:event_classes}) were analyzed: pick and place (object handling parent class), and fetch and deliver (navigation parent class). Pick events involved grasp \& lift motions, while place events involved lowering \& releasing objects. Fetch events consisted of turning and walking to source areas without objects, while deliver events involved carrying objects to target areas. 
Mean durations varied by event type (\autoref{tab:event_stats}): fetching (3.46s) and delivering (3.56s) were longer than picking (1.56s) and placing (1.75s). Object interaction events often occurred in chains of multiple picks or places in succession. All four event classes appeared in each video context, creating 16 possible categorical event distinctions.

\begin{table}
\begin{center}
\caption{Event statistics}
\label{tab:event_stats}
\begin{tabular}{lccc}\toprule
        class & no. of occurrences &	mean duration [s] &	std [s]\\\midrule
        pick	& 82	& 1.56 & 0.79\\
        place	& 104	& 1.75	& 0.57\\
        fetch	& 33	& 3.46	& 0.57\\
        deliver	& 34	& 3.56	& 0.45\\
    \bottomrule
    \end{tabular}
\end{center}
\end{table}

\subsection{Experimental design, participants \& data acquisition}

The experimental design involved two sequences (A and B) of video and resting trials that presented different environments and tasks. Video trials began with a fixation dot followed by a description of the environment (lab 1 or lab 2) and objects (dishes or food), ending with the respective video. In resting trials, participants were shown a fixation dot for 120 seconds, serving as a baseline condition. Each sequence lasted about 20 minutes and included ten video trials and two resting trials with a five-minute break. Sequence A displayed setting plates/cutlery before presenting food/drinks while sequence B showed the reverse order. Each participant was presented each video twice in different sequence orders.

Thirty right-handed, healthy participants (21\female, mean age 23.3 years, SD = 4.54 years) from the campus of the University of Bremen were recruited. The study was approved by the local ethics committee. Participants were informed about MRI safety and the experimental setup and provided written consent to participate. A brief tutorial was given to familiarize with the stimuli and their role in the experiment. 

Participants were situated in a Siemens 3 Tesla MAGNETOM Skyra full body scanner, and after a second tutorial presentation, the experiment was started. The stimuli were presented through a mirror system attached to the head coil. fMRI data were recorded via T2*-weighted multi-band EPI with acceleration factor of 3, TR = 1.1 s, TE = 30 ms, matrix size = 64x64x45 voxels and voxel size = 3x3x3 mm. After the functional MR-recording, a T1-weighted structural scan with a matrix size = 255x265x265 voxels and a voxel size = 1x1x1 mm was conducted.

\subsection{Preprocessing and data analysis}
fMRI data underwent standard preprocessing in SPM12~\citep{SPM12} (MATLAB 2018b~\citep{MATLAB}), including slice-time and motion correction, co-registration, normalization, and 8mm FWHM Gaussian smoothing. Event timing data for pick, place, fetch, and deliver actions were processed for GLM analysis, incorporating motion regressors and a 128s high-pass filter and used for single-subject analyses via contrasting against rest conditions. Resulting contrasts were analyzed in a group-level factorial design with three factors: event class (4 levels), environmental context (2 levels), and item context (2 levels).
Representational Similarity Analysis (RSA)~\citep{nili_toolbox_2014} was performed on the 16 group-level statistical t-maps using MATLAB's RSA toolbox. The analysis resulted in Representational Dissimilarity Matrices (RDMs) as depicted in \autoref{fig:rdm_hrf60} and \autoref{fig:rdm_hrf54}, with 120 dissimilarity values in the range of [0 1], and including 24 within-class and 96 between-class values. 
GLM and RSA analyses were repeated across multiple HRF-Time-to-Peak (TTP) values (4.0-7.0s, 200ms steps). Class distinctiveness was assessed by comparing between-class and within-class correlation distances, with optimal distinctiveness determined by minimal overall overlap.

The resulting t-maps ($p \leq 0.05$, FWE corrected) were analyzed for spatial activation pattern overlap between parent classes. Only HRF-TTP-distinct overlaps were considered, and results were aligned with the human brainnetome atlas~\citep{fan_human_2016}. For optimal HRF-TTP, a three-ranked conjunction analysis examined common activation patterns across all events, object handling versus navigation contrasts, and individual event classes. Results were thresholded, with higher-rank activations subtracted from lower ranks to identify supplemental activity patterns.

\section{Results}
\subsection{Optimal HRF-TTP for categorical clustering}

\begin{figure}
    \centering
    \fbox{\includegraphics[width=.75\columnwidth]{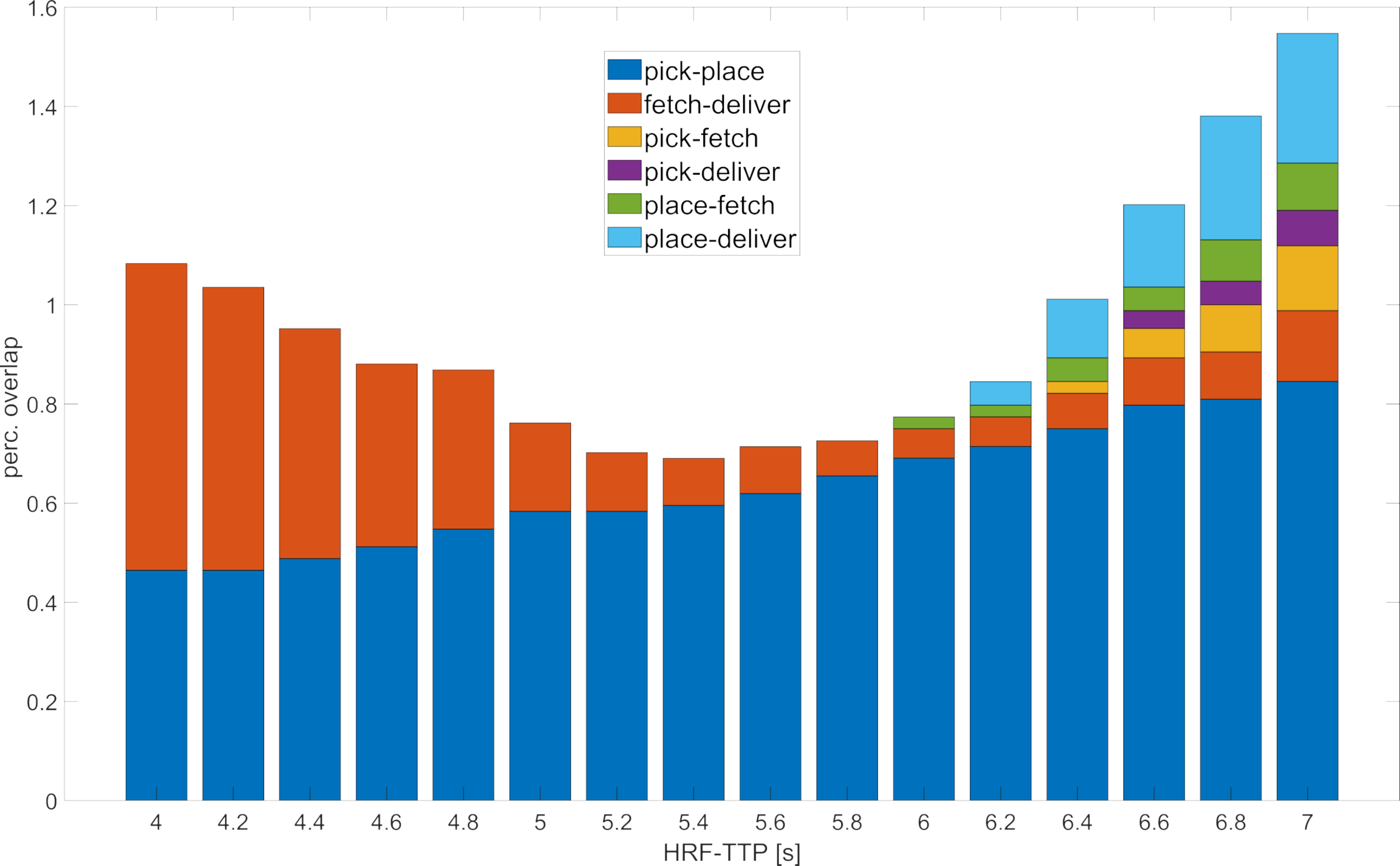}}
    \caption{Cumulative bar graphs of between- and within-class overlap. Bars depict the percentage of overlap between instances of respective between-class correlation distances and  within-class values for all HRF-TTP settings. Results are scaled between [0 1] for each category.}
    \label{fig:overlap}
\end{figure}

The results of the calculation of HRF-TTP values for optimal class distinctiveness are depicted as a bar graph in \autoref{fig:overlap}. The lowest cumulative values were found for HRF-TTP = 5.4s with 9.52\%  of fetch-deliver and 59.52\% of pick-place distances lower than all within-class distances. For the default HRF-TTP of 6.0s, 5.95\% of fetch-deliver, 69.05\% of pick-place, and 2.38\% of place-fetch distances were lower than all within-class distances.
In general, distances between events of pick- and place classes, both belonging to the parent class of object handling, were lowest in relation to within-class distances, and most overlap was found here but a clear trend towards less overlap for smaller HRF-TTP values. The relative distances between the navigation classes of fetch \& deliver exhibited a near opposite pattern with a pronounced rise in overlap for HRF-TTP values below 6.0s. For HRF-TTP values above 6.0s, the within- to between-class distance overlap for fetch \& deliver had a small increase. For these larger HRF-TTP values, the overlap to events belonging to different parent classes became more pronounced, indicating a higher breakdown of class differentiation.

\begin{figure}
    \centering
	\subfloat[RDM: TTP = 6.0 sec]{
        \includegraphics[width=0.35\columnwidth]{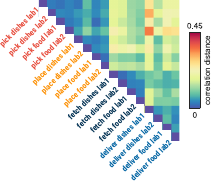}
        \label{fig:rdm_hrf60}
        }
    \hspace{3mm}
    \subfloat[RDM: TTP = 5.4 sec]{
        \includegraphics[width=0.35\columnwidth]{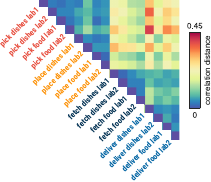}
        \label{fig:rdm_hrf54}
        }\\
        \vspace{.25cm}
    \subfloat[dendrogram: TTP = 6.0 sec]{
        \includegraphics[width=0.47\columnwidth]{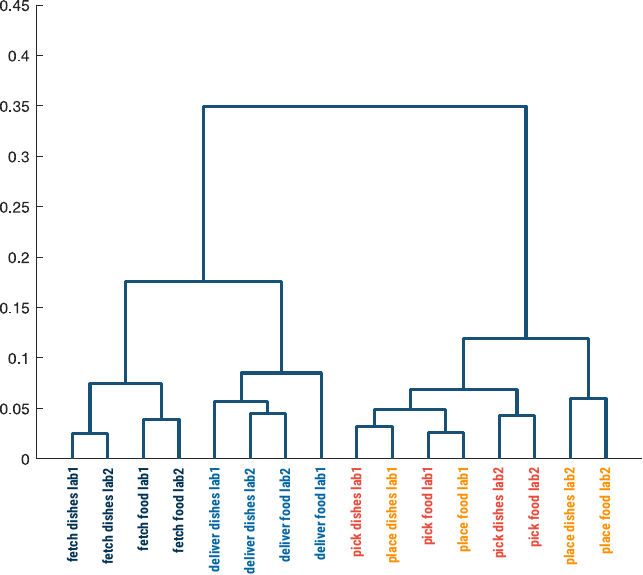}
        \label{fig:dendro_hrf60}
        }
    \subfloat[dendrogram: TTP = 5.4 sec]{
        \includegraphics[width=0.47\columnwidth]{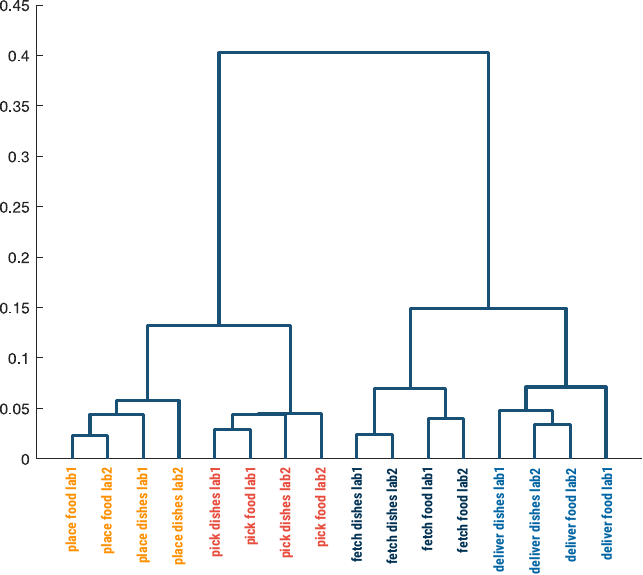}
        \label{fig:dendro_hrf54}
        }\\
	\caption{RDMs and resulting dendrograms from hierarchical clustering with optimal and default HRF-TTP. RDM colors generated via linspecer colormap~\cite{lansey_beautiful_2023}.}
    \label{fig:results_dendro}
\end{figure}

The RDMs and the resulting dendrograms for standard- and optimal HRF-TTP values are depicted in \autoref{fig:results_dendro}. While visual inspection indicated differences in correlation distance between both RDMs, the resulting dendrograms substantiated the impact on categorical clustering quality.
For both HRF-TTP of 6.0s and 5.4s, a grouping of events into the parent classes of object handling and navigation was successful, however with a higher cluster distance for TTP =  5.4s. For the parent class of navigation, the grouping of events into fetch and deliver classes was also accomplished for both HRF-TTP values. For HRF-TTP of 6.0s, the grouping of object handling events into pick and place classes was unsuccessful. This result was fixed for the optimal HRF-TTP value of 5.4s.

\subsection{HRF-TTP related overlap in activation maps}

\begin{figure}
\centering
\begin{subfigure}{0.45\columnwidth}
\centering
\fbox{\includegraphics[width=.9\columnwidth]{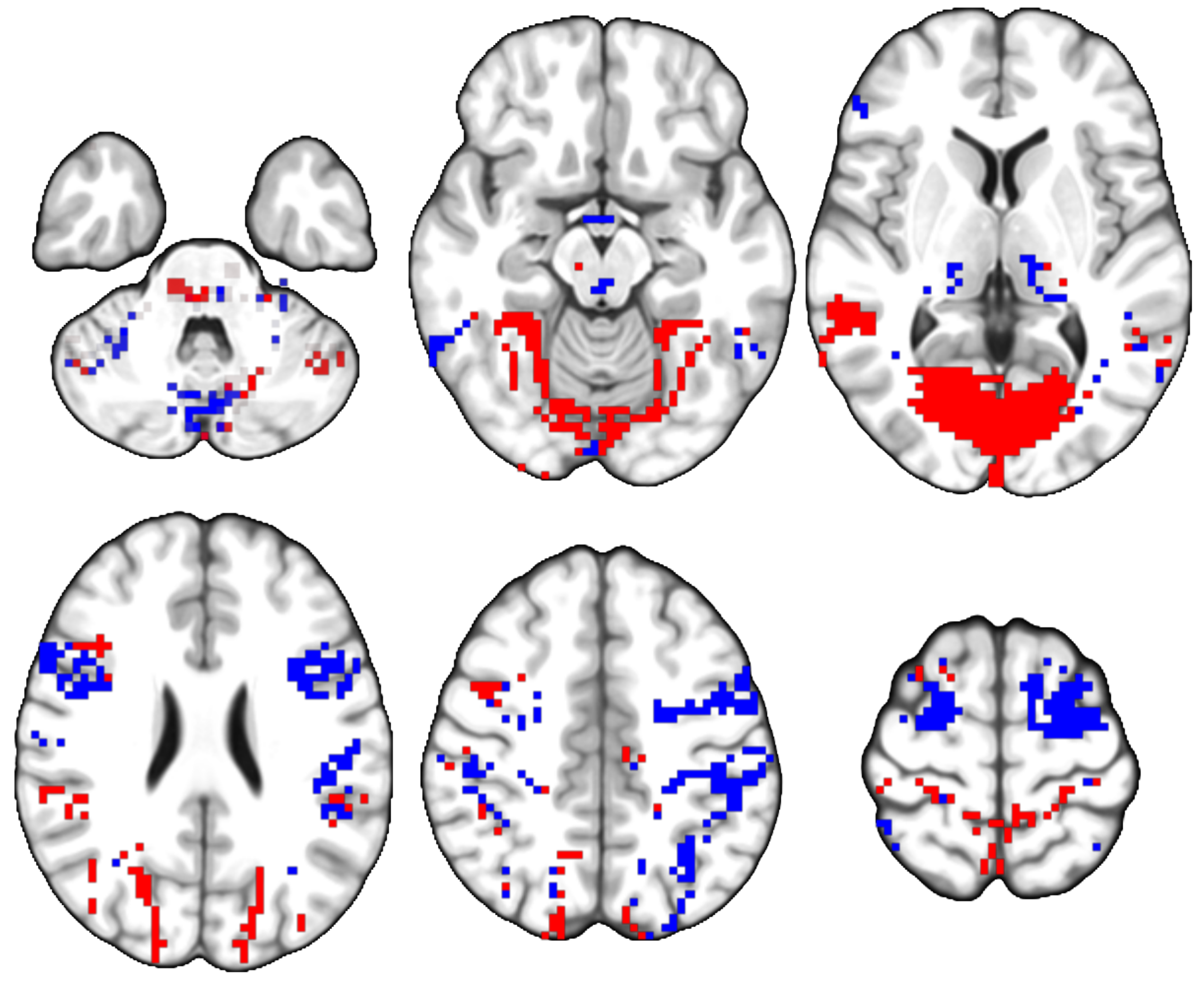}}
\caption{pick - place}
\label{fig:overlap_obj}
\end{subfigure}%
\hfil
\centering
\begin{subfigure}{0.45\columnwidth}
\centering
\fbox{\includegraphics[width=.9\columnwidth]{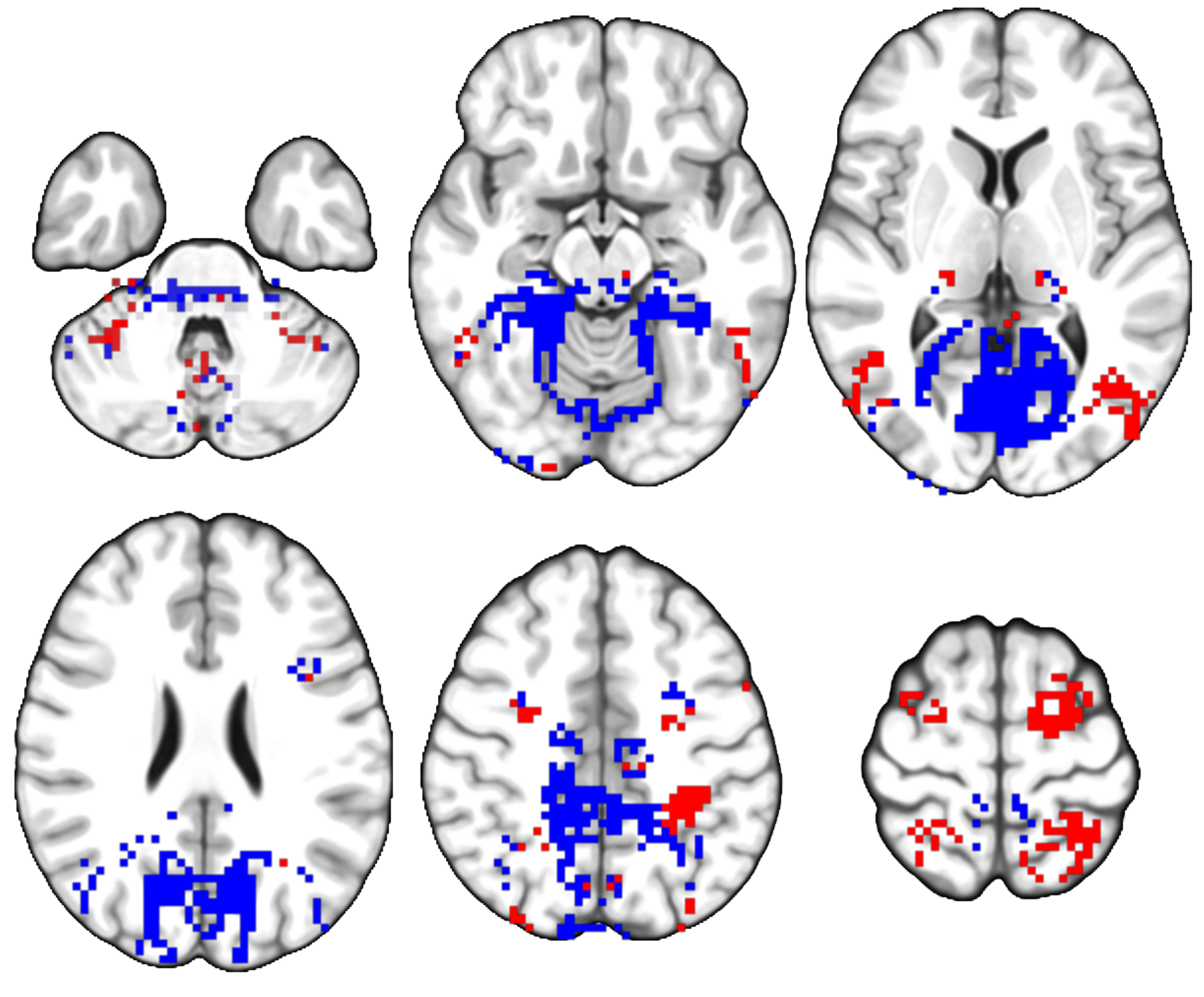}}
\caption{fetch - deliver}
\label{fig:overlap_navi}
\end{subfigure}%
\caption{Areas of distinct overlap between pick - place and fetch - deliver events in activation maps of semantic event representation against baseline at $p \leq 0.05$, FWE corrected. $HRF-TTP = 5.4s$ [blue], $HRF-TTP = 6.0s$ [red]}
\label{fig:overlap_overview}
\end{figure}

HRF-TTP related overlaps in brain activation between pick \& place are listed as number of voxels per brain area in~\autoref{tab:voxel_overlap_detail}, columns 3 \& 4 and depicted via activation maps in figure~\autoref{fig:overlap_obj}.
Distinct overlap of brain activity between pick \& place events for HRF-TTP of 5.4s, as depicted in blue, showed its most prominent clusters in the frontal lobe areas, such as the precentral gyrus, and the parietal lobe with a focus on the inferior parietal lobule. 
For HRF-TTP 6s, as depicted in red, overlap between pick \& place events showed its largest clusters in the occipital lobe, mainly the medioventral occipital cortex. Clusters were also present in the parietal lobe and in the fusiform gyrus of the temporal lobe.

Overlap for fetch \& deliver are listed as number of voxels per brain area in~\autoref{tab:voxel_overlap_detail}, columns 5 \& 6 and via activation maps in figure~\autoref{fig:overlap_navi}.
Overlap between fetch \& deliver events for HRF-TTP of 5.4s, as depicted in blue, showed its largest extent in the occipital lobe, mainly in the medioventral occipital cortex. It was also present in the parietal lobe with a focus on the precuneus. Smaller clusters were found in the temporal-, frontal lobe, and cingulate gyrus.
Overlap between fetch \& deliver events for HRF-TTP of 6.0s, as depicted in red, was largest in the parietal lobe, mainly comprasing the superior parietal lobule. It was further present in the frontal lobe and in the occipital lobe in the lateral occipital cortex. Additional overlap was mainly present in the temporal lobe.
For all event classes and TTP values, there was an additional overlap in the cerebellum.

\begin{table}[t]
\begin{center}
\caption{Voxel count of areas of distinct overlap between pick - place and fetch - deliver events, per brain area\tablefootnote{
SFG = superior frontal gyrus, 
MFG = middle frontal gyrus,
IFG = inferior frontal gyrus,
OrG = orbital gyrus,
PrG = precentral gyrus,
PCL = paracentral lobule,
STG = superior temporal gyrus,
MTG = middle temporal gyrus,
ITG = inferior temporal gyrus,
FuG = fusiform gyrus,
PhG = parahippocampal gyrus,
pSTS = posterior superior temporal sulcus,
SPL = superior parietal lobule,
IPL = inferior parietal lobule,
PCun =	precuneus,
PoG = postcentral gyrus,
INS = insular gyrus,
CG = cingulate gyrus,
MVOcC = medioventral occipital cortex,
LOcC = lateral occipital cortex,
Amyg = amygdala,
Hipp = hippocampus,
BG = basal ganglia,
Tha = thalamus,
Ceb = cerebellum
}
}
\label{tab:voxel_overlap_detail}
\begin{tabular}{
>{\raggedleft\arraybackslash}p{1.1cm}
>{\raggedleft\arraybackslash}p{.9cm}p{.4cm}>{\raggedleft\arraybackslash}p{.8cm}
>{\raggedleft\arraybackslash}p{.8cm}
>{\raggedleft\arraybackslash}p{.8cm}
>{\raggedleft\arraybackslash}p{.8cm}}
\toprule
    \multicolumn{3}{c}{} & \multicolumn{2}{c}{pick \& place} & \multicolumn{2}{c}{fetch \& deliver}
    \\\cmidrule(lr){4-5}\cmidrule(lr){6-7}
    brain & total &&	TTP &	TTP &	TTP &	TTP\\
    area & voxels &&	5.4s &	6.0s &	5.4s &	6.0s\\\midrule
    SFG	&	3315	&&	219	&	14	&	2	&	113	\\
    MFG	&	3585	&&	138	&	44	&	7	&	57	\\
    IFG	&	1312	&&	131	&	22	&	8	&	1	\\
    OrG	&	2484	&&	12	&	6	&	4	&	0	\\
    PrG	&	2144	&&	434	&	29	&	9	&	124	\\
    PCL	&	553	&&	1	&	39	&	119	&	0	\\[.2cm]
    STG	&	2096	&&	9	&	15	&	3	&	0	\\
    MTG	&	1866	&&	10	&	57	&	0	&	19	\\
    ITG	&	1684	&&	55	&	34	&	12	&	93	\\
    FuG	&	1662	&&	12	&	201	&	163	&	34	\\
    PhG	&	508	&&	4	&	2	&	118	&	2	\\
    pSTS	&	398	&&	23	&	68	&	14	&	0	\\[.2cm]
    SPL	&	1404	&&	69	&	37	&	35	&	220	\\
    IPL	&	4038	&&	374	&	219	&	167	&	58	\\
    PCun	&	1740	&&	3	&	259	&	688	&	11	\\
    PoG	&	1680	&&	116	&	74	&	3	&	42	\\[.2cm]
    INS	&	945	&&	0	&	0	&	0	&	0	\\[.2cm]
    CG	&	1655	&&	1	&	12	&	145	&	10	\\[.2cm]
    MVOcC	&	2199	&&	0	&	823	&	1103	&	0	\\
    LOcC	&	2850	&&	11	&	131	&	200	&	183	\\[.2cm]
    Amyg	&	176	&&	1	&	4	&	0	&	0	\\
    Hipp	&	671	&&	6	&	23	&	46	&	4	\\
    BG	&	1454	&&	1	&	4	&	0	&	0	\\
    Tha	&	962	&&	49	&	14	&	20	&	20	\\[.2cm]
    Ceb	&	6135	&&	216	&	341	&	374	&	199	\\[.2cm]
    \textbf{sum}	& \textbf{47516}	&& \textbf{1895}	& \textbf{2472}	& \textbf{3240}	& \textbf{1190}\\\bottomrule
\end{tabular}
\end{center}
\end{table}

\subsection{Activation maps for optimal HRF-TTP values}

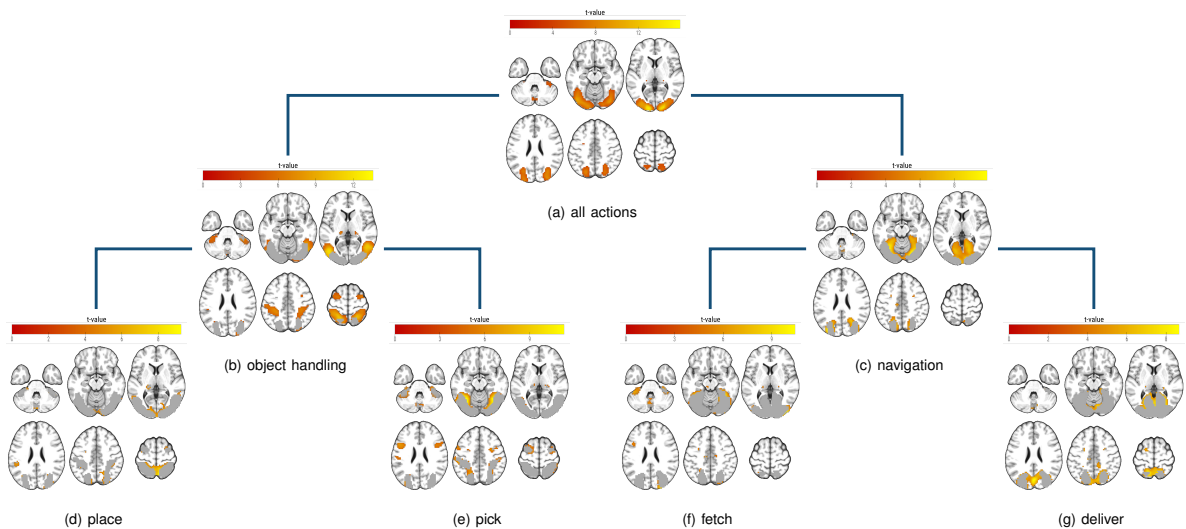
\begin{figure*}[t]
    \centering
    \begin{tcolorbox}[sharp corners,colback=white,frame empty, boxsep=1pt,left=0pt,right=0pt,top=8pt,bottom=8pt]
    \resizebox{\textwidth}{!}{
        \begin{tikzpicture}

    \node (img_cAll) {\includegraphics[width=0.2\columnwidth]{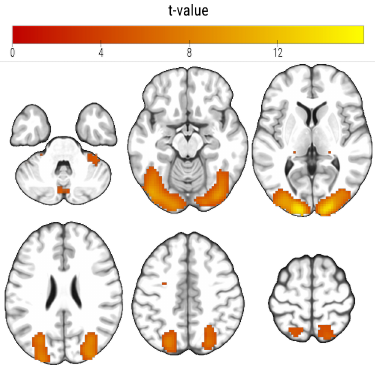}};
    \node[text width=6cm,align=center,anchor=north] (txt_cAll) at ([yshift=-.0mm]img_cAll.south) {\subcaption{all actions \label{fig:conj_all_hrf54}}};

    \node[left=2.25cm of img_cAll, yshift=-3cm] (img_cObj) {\includegraphics[width=0.2\columnwidth]{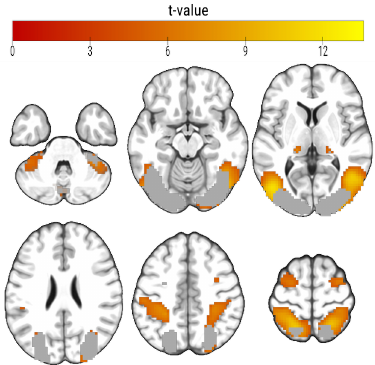}};
    \node[text width=6cm,align=center,anchor=north] (txt_cObj) at ([yshift=-.0mm]img_cObj.south) {\subcaption{object handling \label{fig:masked_conj_all_x_conj_obj_hrf54}}};

    \node[right=2.25cm of img_cAll, yshift=-3cm] (img_cNav) {\includegraphics[width=0.2\columnwidth]{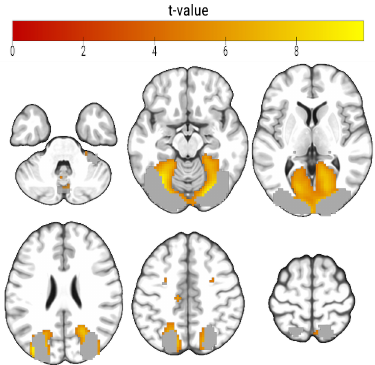}};
    \node[text width=6cm,align=center,anchor=north] (txt_cNav) at ([yshift=-.0mm]img_cNav.south) {\subcaption{navigation \label{fig:masked_conj_all_x_conj_nav_hrf54}}};

    \node[left=0cm of img_cObj, yshift=-3cm] (img_cPlace) {\includegraphics[width=0.2\columnwidth]{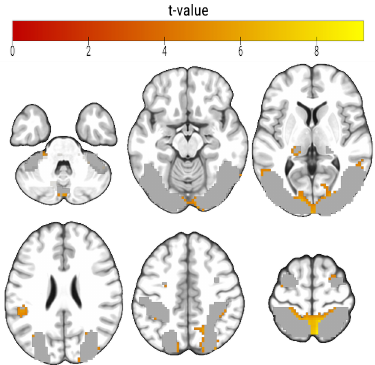}};
    \node[text width=6cm,align=center,anchor=north] (txt_cPlace) at ([yshift=-.0mm]img_cPlace.south) {\subcaption{place \label{fig:masked_conj_obj_x_conj_place_hrf54}}};

    \node[right=-0cm of img_cObj, yshift=-3cm] (img_cPick) {\includegraphics[width=0.2\columnwidth]{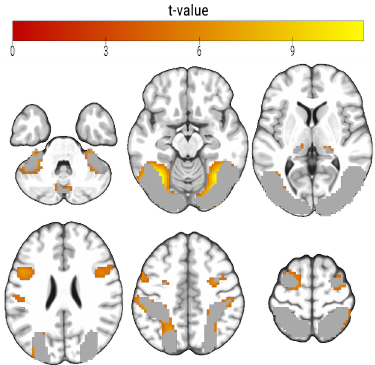}};
    \node[text width=6cm,align=center,anchor=north] (txt_cPick) at ([yshift=-.0mm]img_cPick.south) {\subcaption{pick \label{fig:masked_conj_obj_x_conj_pick_hrf54}}};

    \node[left=-0cm of img_cNav, yshift=-3cm] (img_cFetch) {\includegraphics[width=0.2\columnwidth]{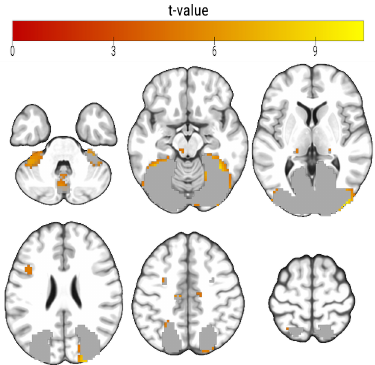}};
    \node[text width=6cm,align=center,anchor=north] (txt_cFetch) at ([yshift=-.0mm]img_cFetch.south) {\subcaption{fetch \label{fig:masked_conj_nav_x_conj_fetch_hrf54}}};

    \node[right=-0cm of img_cNav, yshift=-3cm] (img_cDeliver) {\includegraphics[width=0.2\columnwidth]{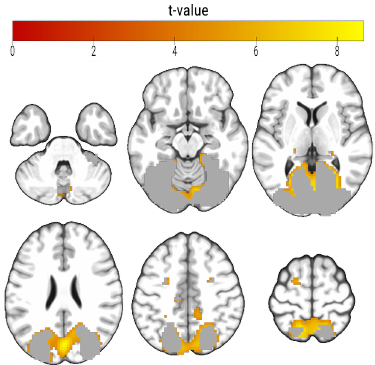}};
    \node[text width=6cm,align=center,anchor=north] (txt_cAll) at ([yshift=-.0mm]img_cDeliver.south) {\subcaption{deliver \label{fig:masked_conj_nav_x_conj_deliver_hrf54}}};    

    \draw[-, ultra thick,easeblue] (img_cAll.west) -| (img_cObj.north);
    \draw[-, ultra thick,easeblue] (img_cAll.east) -| (img_cNav.north);
    \draw[-, ultra thick,easeblue] (img_cObj.west) -| (img_cPlace.north);
    \draw[-, ultra thick,easeblue] (img_cObj.east) -| (img_cPick.north);
    \draw[-, ultra thick,easeblue] (img_cNav.west) -| (img_cFetch.north);
    \draw[-, ultra thick,easeblue] (img_cNav.east) -| (img_cDeliver.north);
    
    \end{tikzpicture}
        }
    \end{tcolorbox}
    \caption{Hierarchical classification of activation maps for contrasts of semantic event presentations greater baseline for $HRF-TTP = 5.4s$, at $p \leq 0.05$, FWE, based on RDM clustering}
	  \label{fig:results_patterns_hrf54}

\end{figure*}

Activation maps of conjunction analyses are shown in~\autoref{fig:results_patterns_hrf54}. Voxel sizes for activation patterns are listed in~\autoref{tab:voxel_opt_detail}. For all conjunction analyses, additional large clusters were shown in the cerebellum.
For the conjunction analysis between all 16 event types (\autoref{fig:conj_all_hrf54}), shared activity was mainly found in lateral occipital cortex extending into the medioventral occipital cortex. It also encompassed activity in the superior- and inferior parietal gyrus and precuneus as well as temporal lobe activity in the precuneus. Smaller clusters were also found in the hippocampus and thalamus.

Supplemental activity for the conjunction analysis of all events of the object handling parent class (\autoref{fig:masked_conj_all_x_conj_obj_hrf54}) showed its largest clusters distributed over the parietal lobe. Large clusters were also found in the middle- and inferior temporal and fusiform gyrus as well as superior- and middle frontal and precentral gyrus and the lateral occipital cortex. A smaller cluster was found in the thalamus.
The conjunction analysis of all eight navigation events (\autoref{fig:masked_conj_all_x_conj_nav_hrf54}) had the most prominent supplemental activity in the medioventral occipital lobe. Other large clusters were shown in a different configuration in the parietal lobe and in the fusiform gyrus of the temporal lobe, while frontal lobe activity was less pronounced. Additional activity was found in the cingulate gyrus and the hippocampus.

At the level of class specific events, placing (\autoref{fig:masked_conj_obj_x_conj_place_hrf54}) resulted in an activity pattern with largest cluster size in the medioventral- and to a lesser extend lateral occipital cortex and in the the parietal lobe with a focus on the precuneus. Smaller clusters were found in the frontal- and temporal lobes and the thalamus.
Pick events (\autoref{fig:masked_conj_obj_x_conj_pick_hrf54}), resulted in largest clusters in the parietal lobe with a different distribution, the temporal lobe with a focus on the fusiform gyrus, and large frontal lobe clusters with a focus on the precentral gyrus. Additional notable clusters were shown in the  occipital lobe and thalamus.
Fetching events (\autoref{fig:masked_conj_nav_x_conj_fetch_hrf54}), resulted in supplemental activity mainly in the lateral occipital cortex and the fusiform gyrus of the temporal lobe. Clusters of smaller extend were found in the parietal- and frontal lobes, hippocampus, and thalamus.
Supplemental activity for delivery events (\autoref{fig:masked_conj_nav_x_conj_deliver_hrf54}) was larger and with a main distribution over the medioventral occipital cortex and smaller clusters in the lateral occipital cortex. In the parietal lobe, a major pattern was found with a focus on the precuneus.

\begin{table*}[h]
\begin{center}
\caption{Voxel count of of activation maps for contrasts of semantic event presentations greater baseline}
\label{tab:voxel_opt_detail}
\begin{tabular}{>{\raggedleft\arraybackslash}p{1.1cm}>{\raggedleft\arraybackslash}p{.9cm}p{.4cm}>{\raggedleft\arraybackslash}p{1cm}p{.4cm}>{\raggedleft\arraybackslash}p{1.2cm}
>{\raggedleft\arraybackslash}p{1cm}p{.4cm}>{\raggedleft\arraybackslash}p{1cm}
>{\raggedleft\arraybackslash}p{1cm}
>{\raggedleft\arraybackslash}p{1cm}
>{\raggedleft\arraybackslash}p{1cm}}
\toprule
    brain & total &&	all  &&	object  &	navi- &&	&	&	& \\
    area & voxels &&	actions &&	handling  &	gation &&   place &	pick    &	fetch	& deliver\\\midrule
	SFG	&	3315	&&	0	&&	114	&	1	&&	17	&	56	&	0	&	58	\\
	MFG	&	3585	&&	12	&&	102	&	7	&&	2	&	76	&	2	&	51	\\
	IFG	&	1312	&&	0	&&	0	&	0	&&	0	&	98	&	15	&	0	\\
	OrG	&	2484	&&	0	&&	0	&	0	&&	6	&	0	&	0	&	0	\\
	PrG	&	2144	&&	0	&&	103	&	1	&&	15	&	272	&	10	&	10	\\
	PCL	&	553	&&	0	&&	0	&	20	&&	45	&	0	&	14	&	13	\\[.2cm]
	STG	&	2096	&&	0	&&	0	&	0	&&	2	&	0	&	0	&	0	\\
	MTG	&	1866	&&	0	&&	166	&	0	&&	13	&	2	&	0	&	0	\\
	ITG	&	1684	&&	9	&&	303	&	0	&&	9	&	46	&	15	&	2	\\
	FuG	&	1662	&&	432	&&	143	&	366	&&	0	&	335	&	233	&	7	\\
	PhG	&	508	&&	0	&&	0	&	53	&&	0	&	0	&	24	&	15	\\
	pSTS	&	398	&&	0	&&	47	&	0	&&	38	&	0	&	0	&	0	\\[.2cm]
	SPL	&	1404	&&	536	&&	651	&	76	&&	98	&	39	&	11	&	163	\\
	IPL	&	4038	&&	351	&&	412	&	106	&&	69	&	198	&	43	&	56	\\
	PCun	&	1740	&&	40	&&	97	&	363	&&	271	&	0	&	28	&	503	\\
	PoG	&	1680	&&	0	&&	307	&	0	&&	91	&	65	&	0	&	0	\\[.2cm]
	INS	&	945	&&	0	&&	0	&	0	&&	0	&	0	&	0	&	0	\\[.2cm]
	CG	&	1655	&&	0	&&	0	&	75	&&	0	&	0	&	4	&	42	\\[.2cm]
	MVOcC	&	2199	&&	321	&&	0	&	963	&&	123	&	67	&	1	&	735	\\
	LOcC	&	2850	&&	1994	&&	472	&	74	&&	46	&	16	&	181	&	92	\\[.2cm]
	Amyg	&	176	&&	0	&&	0	&	0	&&	0	&	0	&	0	&	0	\\
	Hipp	&	671	&&	52	&&	5	&	16	&&	3	&	2	&	16	&	3	\\
	BG	&	1454	&&	0	&&	0	&	0	&&	0	&	0	&	0	&	0	\\
	Tha	&	962	&&	47	&&	48	&	0	&&	24	&	16	&	19	&	9	\\[.2cm]
	Ceb	&	6135	&&	382	&&	418	&	278	&&	162	&	391	&	317	&	190	\\[.2cm]
	\textbf{sum}&	\textbf{47516}	&&	\textbf{4176}	&&	\textbf{3388}	&	\textbf{2399}	&&	\textbf{1034}	&	\textbf{1679}	&	\textbf{933}	&	\textbf{1949}\\\bottomrule

\end{tabular}
\end{center}
\end{table*}

\section{Discussion}
The present approach of clustering brain activation patterns from optimal HRF-TTP resulted in a distinct hierarchical structure for the respective event categories. The analysis of underlying functional neuroanatomy substantiated these findings with a common network active for all events and distinct supplemental areas for the parent classes of object handling and navigation with successively less differences for each of the two semantic classes within each parent classes. Furthermore, these brain patterns corresponded to hypotheses concerning their involvement in real-world environmental interaction.

Starting with brain activity observed for optimal clustering HRT-TTP value, the common network active for all actions had its most prominent pattern in the occipital lobe with brain activation related to early visual perception and classification of motions, objects and scenes but with lower involvement of whole-scene related brain areas of the medioventral occipital cortex \citep{dupont_kinetic_1997,malikovic_cytoarchitecture_2016,takahashi_pure_2002}. Brain areas in the parietal- and temporal lobes such as BA~7, BA~37, and BA~39 were found to play a distinctive role in downstream processing of visual stimuli in relation to internal spatial representation, semantic search processes, object identification, fact retrieval as well as memory allocation and planning and suppression of limb movement \citep{ardila_language_2015,culham_human_2006,seghier_angular_2013} with BA~39 acting as multi-modal integration center \citep{binder_where_2009}. 

This basic network for processing visuospatial information was complemented depending on the semantic identity of the presented events and their respective processing need or cognitive load. On the rank of the parent classes, navigation events resulted in supplemental activation with largest extend in occipital areas involved in whole-scene and landscape processing as well as topographical learning and spatial orientation together with brain areas involving inferior temporal (fusiform) brain areas (BA~37) and the parahippocampus as well as the cingulate gyrus (BA~23) \citep{janzen_selective_2004,takahashi_pure_2002,vogt_cytology_2006}. In the parietal lobe, further recruitment involved brain areas related to multi-modal integration, movement planning and the suppression of active action plans.
For fetch events, supplemental activation was most prominent in the lateral occipital cortex in areas processing objects and object-related motion, while additional areas in the temporal cortex were involved in object recognition and topographical processing. In the frontal lobe, additional activity was found in the premotor cortex involved in formulation of motion plans and mapping of available actions to perceived objects with inclusion of BA~44/45 that might play a role in semantic classification processes \citep{maranesi_cortical_2014,sakai_prefrontal_2006}. It can thus be hypothesized that fetching events alter the navigation network in anticipation of subsequently identifying and retrieving objects from the source area.
For delivery events, the navigation network had a larger supplemental extension into occipital lobe areas involved in whole scene processing as well as parietal areas such as BA~5 and BA~7 involved in planning and execution of upper limb movements \citep{culham_role_2006} and BA~23 involved in spatial orientation. Frontal lobe activation was also higher with more extensive engagement of the premotor cortex but no involvement of the inferior frontal gyrus (BA~44/45), hinting at a far higher overall engagement in object interaction and general spatial orientation during this navigation class but less need of object recognition and classification systems. Since one or more objects were already held during these events, larger focus on the spatial and motor related aspect can be explained with the need for preparation of the subsequent placing event(s).

In contrast to navigation, supplemental network recruitment for object handling events involved no whole-scene or spatial orientation related areas but seemed to prepare the network to act in a more object focused manner. In the occipital lobe, additional activity was found in areas related to object motion processing, while main supplemental network activity was present as wide ranging extension of parietal and temporal brain areas involved in the processes multi-modal integration semantic search, object identification, planning and suppression of limb movement \citep{mackenzie_human_2016} now extending into of areas involved in processing of sequential event timings \citep{guidali_keeping_2019} and parts of the somatosensory cortex. For these events, frontal lobe activity was also more pronounced with a focus on the premotor cortex and its role in action planning and mapping.
Additional activation for pick events was found as spatial extension within object- and self-movement related areas of the temporal and parietal cortices that were already present in the more generalized object handling network. Additional clusters in the frontal lobe, encompassing the premotor cortex with inclusion of BA~44/45 as well as parts of the primary motor cortex illustrate the increased need for action classification, planning and executive functions during pick events. 

HRF-TTP related variability of GLM results between standard clustering HRF-TTP of 6.0s and the optimal value of 5.4s, as analyzed through RSA based hierarchical clustering, manifested in higher class consistency for the optimal value with lower within-class correlation distances when compared with distances between events of sibling classes for the parent class of object interaction. The opposite pattern was observed with a lesser extend for the parent class of navigation that did not lead the to a breakdown of the hierarchical clustering of the fetch \& deliver sibling classes. 

The additional analysis of voxel-wise overlap in thresholded t-maps maps provided supplemental information concerning possible factors from functional neuroanatomy. For the standard HRF-TTP value of 6.0s, it was observed that areas of voxel overlap between pick and place events encompassed, among others, areas such as the medioventral occipital cortex and precuneus that were mainly present for navigation events in the conjunction analysis with optimal SOMA-congruent clustering value of 5.4s. For this optimal HRF-TTP value, overlap between pick \& place events focused towards areas in the frontal- and parietal cortex, associated with object identification and interaction. Overlap between fetch \& deliver events at the standard HRF-TTP value of 6.0s included areas more closely associated with object interaction events in the conjunction analysis while the optimal value of 5.4s led to overlap in navigation related areas.
Due to the mean event durations of 1.56s and 1.75s for pick and place, respectively 3.46s and 3.56s for fetch and deliver events, it can be speculated that the convolution of event timings with the longer standard HRF-TTP would misalign the GLM conditions to shift into the subsequent parent class. Since events of the object interaction parent class would generally lead to ones of the navigation class and vice versa, activation maps thus incorporated both object interaction and navigation characteristics, leading to a less 'pure' class definition. This was also expressed through the higher cluster distance between object interaction navigation classes in the hierarchical clustering at optimal HRF-TTP value. The difference would not affect events of the navigation class as much, since the relative shift was not as pronounced due to their longer mean duration.

The activation maps from the voxel-wise overlap analysis thus provided qualitative descriptions for differences in HRF-TTP related clustering on the level of parent classes. For the differences in clustering on the lowest level, i.e., what is the functional reason behind the non-alignment of the hierarchical clustering with the SOMA ontology for standard HRF-TTP for the classes of pick \& place, it corroborates the finding from the RSA clustering analysis by showing higher numbers of overlapping voxels for standard HRF-TTP between pick \& place events while exhibiting the opposite characteristics between fetch \& deliver events. 
For potential functional factors, an in-depth analysis of the neuronal patterns corresponding each event condition would thus be a necessary next step. 

Furthermore, while observed HRF-TTP related differences in automatic clustering are in parts based on the statistical model, with object interaction events exhibiting shorter lengths and a more burst-like timing compared to navigation events, it can be hypothesized that a temporal factor in brain dynamics is revealed via changes in HRF-TTP. Temporo-spatial dynamics could thereby change the neuronal pattern configuration from 'this is a pick event no matter the context' to 'this is an object handling event involving a certain item or environment'. An ambiguity relating to variation of HRF-TTP values might thus not necessarily be an error of the statistical model but important part of temporo-spatial attributes of the human neuronal network that could be probed with changes in stimulus convolution.

\section{Conclusion}
The present data robustly support our hypothesis, that brain activity can be categorized into semantic event classes through General Linear Model (GLM) analyses and Representational Similarity Analysis (RSA), and that the resulting activity patterns would closely resemble those hypothesized in real-world interaction.
By adjusting HRF parameters, we aligned these patterns more closely with the SOMA ontology, underscoring the ongoing need for refinement. These findings not only deepen our understanding of how the brain organizes real-world interactions but also highlight the
necessity for evolving neuroimaging techniques to more accurately mirror the dynamic and context-sensitive brain activity for the robot ontologies. This alignment supports the validity and demonstrates potential refinements of the SOMA ontology.   

\section{Acknowledgements}
 The research reported in this paper has been supported by the German Research Foundation DFG, as part of Collaborative Research Center 1320 EASE - \emph{Everyday Activity Science and Engineering}.

\bibliographystyle{ccn_style}
\bibliography{literature}

\end{document}